\documentclass[aps, prd, showkeys, showpacs ]{revtex4}

\usepackage{amsmath,amsfonts,amscd,amssymb,epsf, multirow}
\usepackage[small]{caption}

\newcommand{\pa}{\partial}

\newcommand{\vep}{\varepsilon}

\begin{document}

\title{ Sphere-plate Casimir interaction    in $\boldsymbol{(D+1)}$-dimensional spacetime}

\author{L. P. Teo}
 \email{LeePeng.Teo@nottingham.edu.my}
 \affiliation{Department of Applied Mathematics, Faculty of Engineering, University of Nottingham Malaysia Campus, Jalan Broga, 43500, Semenyih, Selangor Darul Ehsan, Malaysia.}
\begin{abstract}
In this paper, we derive the formula for the Casimir interaction energy between a sphere and a plate in $(D+1)$-dimensional Minkowski spacetime. It is assumed that the scalar field satisfies the Dirichlet or Neumann boundary conditions on the sphere and the plate. As in the $D=3$ case, the formula is of TGTG type. One of our main contributions is  deriving the translation matrices which express the change of bases between plane waves and spherical waves for general $D$. Using orthogonality of Gegenbauer polynomials, it turns out that the final TGTG formula for the Casimir interaction energy can be simplified to one that is similar to the $D=3$ case. To illustrate the application of the formula, both large separation and small separation asymptotic behaviors of the Casimir interaction energy are computed. The large separation leading term is proportional to $L^{-D+1}$ if the sphere is imposed with Dirichlet boundary condition, and to $L^{-D-1}$ if the sphere is imposed with Neumann boundary condition, where $L$ is distance from the center of the sphere to the plane. For the small separation asymptotic behavior, it is shown that the leading term is equal to the one obtained using proximity force approximation. The next-to-leading order term is also computed using perturbation method. It is shown that when the space dimension $D$ is larger than 5, the next-to-leading order has sign opposite to the leading order term. Moreover, the ratio of the next-to-leading order term to the leading order term is linear in $D$, indicating a larger correction at higher dimensions.
\end{abstract}

\maketitle
\section{Introduction}
The study of Casimir effect has been inspired by the advancement of a number of theoretical and experimental disciplines in physics. After its experimental confirmation in the end of last century, Casimir effect has become a non-negligible  phenomena that draws more and more attention (see e.g., the review \cite{8}).

For more than thirty years, researchers have been considering Casimir effect in $(D+1)$-dimensional spacetime, see for example, \cite{9,10,11,25}. Since exploring physics in higher dimensional spacetime has become a central theme in theoretical physics,  studying Casimir effect in higher dimensional spacetime has become a norm. A lots of work have been done in the last ten years. Most of the works considered the geometry of two parallel plates or branes \cite{12}, and there are also a few works on concentric spheres \cite{13,14,15,16}.

In this paper, we want to take the first step to investigate the Casimir interaction between any two objects in higher dimensional spacetime. We consider massless scalar field for the sphere-plate geometry.

Sphere-plate geometry is the most popularly studied nontrivial configurations for Casimir effect.
In the past ten years, a lots of studies have been devoted to compute the Casimir interaction between a sphere and a plate in $(3+1)$-dimensional spacetime. One of the most powerful methods to derive the exact expression for the Casimir interaction energy is the multiple scattering approach in different disguises \cite{1,2,26,3,4,27,28}. This method has since been generalized to compute the Casimir interaction energy between any two objects in $(3+1)$-dimensional spacetime \cite{5,6}. It was shown that the Casimir interaction energy between two objects can be written in the functional form
\begin{equation}\label{eq12_2_1}
E_{\text{Cas}}=\frac{\hbar}{2\pi}\int_0^{\infty} d\xi \ln\det\left(1-\mathbb{T}_1\mathbb{U}_{12}\mathbb{T}_2\mathbb{U}_{21}\right)
\end{equation}where $\xi$ is the imaginary frequency, $\mathbb{T}_1$ and $\mathbb{T}_2$ are the Lippmann-Schwinger T-operators of the two objects which are related to  the scattering matrices of the objects, $\mathbb{U}_{ij}$ is the translation matrix which changes the wave basis centered at object $i$ to the wave basis centered at object $j$.

Eq. \eqref{eq12_2_1} is known as the TGTG formula. In \cite{7}, we have interpreted this formula from the mode summation point of view. Explicit recipes have been given to compute the $\mathbb{T}_i$ and $\mathbb{U}_{ij}$ matrices. From our interpretation, it is easy to see that the TGTG formula can be applied to higher dimensional spacetime, and the prescriptions for computing the $\mathbb{T}_i$ and $\mathbb{U}_{ij}$ matrices remain unchange.

The goal of this paper is to compute the Casimir interaction energy between a sphere and a plate subject to Dirichlet or Neumann boundary conditions in $(D+1)$-dimensional spacetime. The hardest part of the problem is to compute the translation matrices $\mathbb{U}_{ij}$, which is one of the main contributions of this paper.

After deriving the explicit formula for the Casimir interaction energy, we compute the small separation and large separation behaviors of the Casimir interaction energy as a function of the dimension $D$.
As in the $(3+1)$-dimensional case, computing the large separation behavior is in general straightforward, because its leading behavior is determined by a few terms with lowest wave numbers. However,  computing the small separation behavior is in general very complicated. We employ the method introduced by Bordag in \cite{3} to compute analytically the next-to-leading order of the Casimir interaction energy in the cylinder-plate geometry. This method has   also been employed for other geometries \cite{17,18,19,20,21,22,23,24}.

\section{The Casimir interaction between a sphere and a plate}
In this section, we consider the Casimir interaction energy between a sphere and a plate in $(D+1)$-dimensional Minkowski spacetime equipped with the standard metric
$$ds^2=dt^2-dx_1^2-\ldots-dx_D^2.$$ Since the $D=3$ case has been considered extensively \cite{1,26,3,5,6,7}, in the following, we restrict ourselves to $D\geq 4$, although most of arguments still hold for $D=3$.

Assume that the sphere has radius $a$ and is centred at the origin, whereas the plate is located at $x_1=L$ where $L>a$. In the following, we assume that the plate is of dimension $H\times H\times\ldots\times H$, where $H\gg L$.

To derive the Casimir interaction energy between the sphere and the plate, we use the approach discussed in \cite{7}, where   the TGTG  formula for the Casimir interaction energy
\begin{equation}\label{eq12_3_1}
E_{\text{Cas}}=\frac{\hbar}{2\pi}\int_0^{\infty} d\xi \ln\det\left(1-\mathbb{T}_1\mathbb{U}_{12}\mathbb{T}_2\mathbb{U}_{21}\right)
\end{equation} is derived from the point of view of mode summation approach, and is easy to see to be independent of the spacetime dimension.

First, we need to choose coordinates centered at the plate and the sphere and the find the corresponding  bases of regular and outgoing waves $\varphi^{\text{reg}}$ and $\varphi^{\text{out}}$.

For the plate, we choose rectangular coordinate system centered at $\boldsymbol{L}=(L,0,\ldots,0)$. The scalar field $\varphi$ satisfies the equation
\begin{equation}\label{eq12_3_2}
\left(\frac{\pa^2}{\pa x_1^2}+\ldots+\frac{\pa^2}{\pa x_D^2}\right)\varphi = -\frac{\omega^2}{c^2}\varphi.
\end{equation}
Hence, bases of regular and outgoing plane waves are  parametrized by $\boldsymbol{k}_{\perp}=(k_2, k_3, \ldots, k_D)\in \mathbb{R}^{D-1}$, with
\begin{equation*}
\begin{split}
\varphi^{\text{reg}}_{\boldsymbol{k}_{\perp}}(\boldsymbol{x},k)=& \exp\left(-i\sqrt{k^2-k_{\perp}^2}x_1+i\boldsymbol{k}_{\perp}\cdot\boldsymbol{r}_{\perp}\right),\\
\varphi^{\text{out}}_{\boldsymbol{k}_{\perp}}(\boldsymbol{x},k)=& \exp\left(i\sqrt{k^2-k_{\perp}^2}x_1+i\boldsymbol{k}_{\perp}\cdot\boldsymbol{r}_{\perp}\right)
\end{split}
\end{equation*}Here $\boldsymbol{r}_{\perp}=(x_2,x_3,\ldots,x_D)$, $k_{\perp}=\Vert \boldsymbol{k}_{\perp}\Vert=\sqrt{k_2^2+\ldots+k_D^2}$ and $$k=\frac{\omega}{c}.$$

For the sphere, we choose hyperspherical coordinate system: \begin{align*}
x_1=&r\cos\theta_1\\
x_2=&r\sin\theta_1\cos\theta_2\\
&\vdots\\
x_{D-1}=&r\sin\theta_1\ldots\sin\theta_{D-2}\cos\theta_{D-1}\\
x_D=&r\sin\theta_1\ldots\sin\theta_{D-2}\sin\theta_{D-1}
\end{align*}centered at the origin.
When $\boldsymbol{r}=(x_1,\ldots,x_D)$ ranges over $\mathbb{R}^D$, $r$ ranges from 0 to $\infty$, whereas
\begin{align*}
0\leq \theta_i\leq\pi, \quad i=1,2,\ldots,D-2,
\end{align*} and
$$-\pi\leq\theta_{D-1}\leq\pi.$$In the following, we will denote by $S^{D-1}$ the region
\begin{align*}
0\leq \theta_i\leq\pi, \quad i=1,2,\ldots,D-2;\quad -\pi\leq\theta_{D-1}\leq\pi.
\end{align*}
The volume element $\displaystyle\prod_{i=1}^D dx_i$ is equal to $$r^{D-1}\prod_{i=1}^{D-1}\sin^{D-i-1}\theta_id\theta_i $$ in spherical coordinates. Denote by $d\Omega_{D-1}$ the measure
$$\prod_{i=1}^{D-1}\sin^{D-i-1}\theta_id\theta_i.$$ Then
$$\int_{S^{D-1}}d\Omega_{D-1}=\frac{2\pi^{\frac{D}{2}}}{\Gamma\left(\frac{D}{2}\right)}$$ is the volume of the unit sphere $x_1^2+x_2^2+\ldots+x_{D}^2=1$.

In spherical coordinates, the equation of motion \eqref{eq12_3_2} becomes
\begin{equation*}
\left(\frac{\pa^2}{\pa r^2}+\frac{D-1}{r}\frac{\pa}{\pa r}+\frac{1}{r^2}\sum_{i=1}^{D-1}\frac{1}{\prod_{j=1}^{i-1}\sin^2\theta_j}\left(\frac{\pa^2}{\pa\theta_i^2}
+(D-i-1)\frac{\cos\theta_i}{\sin\theta_i}\frac{\pa}{\pa\theta_i}\right)\right)\varphi=-\frac{\omega^2}{c^2}\varphi.
\end{equation*}
The solutions of this differential equation are parametrized by $\boldsymbol{m}=(m_1,\ldots,m_{D-1})$, with
$$l=m_1\geq m_2\geq \ldots\geq m_{D-2}\geq |m_{D-1}|.$$The regular and outgoing spherical waves are \cite{29,30}:
\begin{equation}\label{eq12_3_6}
\begin{split}
\varphi_{\boldsymbol{m}}^{\text{reg}}(\boldsymbol{x},k)=\mathcal{C}_l^{\text{reg}}C_{\boldsymbol{m}}j_l( k r)\boldsymbol{Y}_{\boldsymbol{m}}(\boldsymbol{\theta}),\\
\varphi_{\boldsymbol{m}}^{\text{out}}(\boldsymbol{x},k)=\mathcal{C}_l^{\text{out}}C_{\boldsymbol{m}}h_l^{(1)}( k r)\boldsymbol{Y}_{\boldsymbol{m}}(\boldsymbol{\theta}),
\end{split}
\end{equation}
where
\begin{align*}
j_l(z)=z^{-\frac{D-2}{2}}J_{l+\frac{D-2}{2}}(z),\hspace{1cm}h_l^{(1)}(z)=z^{-\frac{D-2}{2}}H^{(1)}_{l+\frac{D-2}{2}}(z),
\end{align*}
\begin{align*}
\boldsymbol{Y}_{\boldsymbol{m}}(\boldsymbol{\theta})=e^{im_{D-1}\theta_{D-1}}\prod_{j=1}^{D-2}\sin^{m_{j+1}}\theta_{j}C_{m_j-m_{j+1}}^{m_{j+1}+\frac{D-j-1}{2}}(\cos\theta_j),
\end{align*}$J_{l+\frac{D-2}{2}}(z)$ and $H^{(1)}_{l+\frac{D-2}{2}}(z)$ are Bessel functions, and $C_n^{\nu}(z)$  is a Gegenbauer polynomial  defined by
\begin{align*}
(1-2zt+t^2)^{-\nu}=\sum_{n=0}^{\infty} C_n^{\nu}(z)t^n.
\end{align*} The Gegenbauer polynomials satisfy the orthogonality relation
\begin{align}\label{eq12_3_11}
\int_{-1}^1dx C_n^{\nu}(x)C_m^{\nu}(x)(1-x^2)^{\nu-\frac{1}{2}}=\frac{\pi 2^{1-2\nu}\Gamma(n+2\nu)}{n!(n+\nu)\Gamma(\nu)^2}\delta_{n,m}.
\end{align}
Hence, the hyperspherical harmonics $\boldsymbol{Y}_{\boldsymbol{m}}(\boldsymbol{\theta})$ satisfy the orthogonality condition
\begin{equation*}
\int_{S^{D-1}} d\Omega_{D-1}\boldsymbol{Y}_{\boldsymbol{m}}(\boldsymbol{\theta}) \boldsymbol{Y}_{\boldsymbol{m}'}(\boldsymbol{\theta})^* =\frac{1}{C_{\boldsymbol{m}}^2}\delta_{\boldsymbol{m},\boldsymbol{m}'},
\end{equation*}
where
\begin{align*}
C_{\boldsymbol{m}}=\sqrt{\frac{1}{2\pi^{D-1}}\prod_{j=1}^{D-2}\frac{2^{2m_{j+1}+D-j-2}\Gamma\left(m_{j+1}+\frac{D-j-1}{2}\right)^2\left(m_j+\frac{D-j-1}{2}\right)(m_j-m_{j+1})!}
{\Gamma\left(m_j+m_{j+1}+D-j-1\right)}}.
\end{align*}
The constants $\mathcal{C}_l^{\text{reg}}$ and $\mathcal{C}_l^{\text{out}}$ are defined by
\begin{align*}
\mathcal{C}_l^{\text{reg}}=i^{-l},\quad \mathcal{C}_l^{\text{out}}=\frac{\pi}{2}i^{l+D-1},
\end{align*}  so that
\begin{align*}
\mathcal{C}_l^{\text{reg}}j_l(iz)=z^{-\frac{D-2}{2}}I_{l+\frac{D-2}{2}}(z),\hspace{1cm}\mathcal{C}_l^{\text{out}}h^{(1)}_l(iz)=z^{-\frac{D-2}{2}}K_{l+\frac{D-2}{2}}(z).
\end{align*}

The $\mathbb{T}_1$ and $\mathbb{T}_2$ in the TGTG formula \eqref{eq12_3_1}  are the Lippmann-Schwinger T-operators of the sphere and the plate respectively. As in \cite{7}, it is easy to find that for Dirichlet (D) and Neumann  (N) boundary conditions, they are diagonal in $\boldsymbol{m}$  and $\boldsymbol{k}_{\perp}$  respectively with diagonal elements given respectively by
\begin{equation*}
\begin{split}
&T_{1,\boldsymbol{m}}^{\text{D}}(\kappa)=\frac{I_{l+\frac{D-2}{2}}(\kappa a)}{K_{l+\frac{D-2}{2}}(\kappa  a)},\\
&T_{1,\boldsymbol{m}}^{\text{N}}(\kappa)=\frac{-\frac{D-2}{2}I_{l+\frac{D-2}{2}}(\kappa a)+\kappa aI_{l+\frac{D-2}{2}}'(\kappa a)}
{-\frac{D-2}{2}K_{l+\frac{D-2}{2}}(\kappa a)+\kappa a K_{l+\frac{D-2}{2}}'(\kappa a)},\\
&T_{2,\boldsymbol{k}_{\perp}}^{\text{D}}=1,\quad T_{2,\boldsymbol{k}_{\perp}}^{\text{N}}=-1,\\
\end{split}
\end{equation*}Here $\kappa=\xi/c$ and $k=i\kappa$.

The translation matrix $\mathbb{U}_{12}=\mathbb{V}$ in \eqref{eq12_3_1} is defined by
\begin{equation}\label{eq12_3_5}
\varphi_{\boldsymbol{k}_{\perp}}^{\text{reg}}(\boldsymbol{x}-\boldsymbol{L},k)=\sum_{\boldsymbol{m}}V_{\boldsymbol{m},\boldsymbol{k}_{\perp}}\varphi_{\boldsymbol{m}}^{\text{reg}}(\boldsymbol{x},k),
\end{equation}where the summation over $\boldsymbol{m}$ is
\begin{align*}
\sum_{\boldsymbol{m}}=\sum_{l=0}^{\infty}\sum_{m_2=0}^{l}\sum_{m_3=0}^{m_2}\ldots\sum_{m_{D-2}=0}^{m_{D-3}}\sum_{m_{D-1}=-m_{D-2}}^{m_{D-2}}.
\end{align*}The matrix $\mathbb{U}_{21}=\mathbb{W}$ is defined by
\begin{equation}\label{eq12_3_7}
\varphi_{\boldsymbol{m}}^{\text{out}}(\boldsymbol{x}+\boldsymbol{L},k)=\frac{H^{D-1}}{(2\pi)^{D-1}}\int_{\mathbb{R}^{D-1}}d\boldsymbol{k}_{\perp}
W_{\boldsymbol{k}_{\perp},\boldsymbol{m}}\varphi_{\boldsymbol{k}_{\perp}}^{\text{out}}(\boldsymbol{x},k).
\end{equation}In the following, we will derive the explicit expressions for $V_{\boldsymbol{m},\boldsymbol{k}_{\perp}}$ and $W_{\boldsymbol{k}_{\perp},\boldsymbol{m}}$.

The derivation of $V_{\boldsymbol{m},\boldsymbol{k}_{\perp}}$ is relatively easier.
Express $\boldsymbol{k}=(k_1,k_2,\ldots,k_D)$ in hyperspherical coordinates:
\begin{equation}\label{eq12_3_8}\begin{split}
k_1=&k\cos\theta^k_1,\\
k_2=&k\sin\theta^k_1\cos\theta^k_2,\\
&\vdots\\
k_{D-1}=&k\sin\theta^k_1\ldots\sin\theta^k_{D-2}\cos\theta^k_{D-1},\\
k_{D}=&k\sin\theta^k_1\ldots\sin\theta^k_{D-2}\sin\theta^k_{D-1},\end{split}
\end{equation}
and let $S^{D-1}_k$ be the region
\begin{gather*}
0\leq \theta^k_j\leq \pi,\quad 1\leq j\leq D-2,\quad
-\pi\leq \theta^k_{D-1}\leq\pi.
\end{gather*}
As is shown in \cite{30},
\begin{align}\label{eq12_3_4}
\int_{S^{D-1}_k}d\Omega_{D-1}^k\boldsymbol{Y}_{\boldsymbol{m}}(\boldsymbol{\theta}_k)e^{i\boldsymbol{k}\cdot\boldsymbol{r}}=(2\pi)^{\frac{D}{2}} i^{-l}j_l(kr)\boldsymbol{Y}_{\boldsymbol{m}}(\boldsymbol{\theta}).
\end{align}
This is equivalent to
\begin{align*}
e^{i\boldsymbol{k}\cdot\boldsymbol{r}}= (2\pi)^{\frac{D}{2}} \sum_{\boldsymbol{m}}
i^{-l} C_{\boldsymbol{m}}^2j_l(kr)\boldsymbol{Y}_{\boldsymbol{m}}(\boldsymbol{\theta})\boldsymbol{Y}_{\boldsymbol{m}}(\boldsymbol{\theta}_k)^*.
\end{align*}Comparing this to the definition of $V_{\boldsymbol{m},\boldsymbol{k}_{\perp}}$ \eqref{eq12_3_5} and $\varphi_{\boldsymbol{m}}^{\text{reg}}(\boldsymbol{x},k)$ \eqref{eq12_3_6}, we obtain immediately
\begin{align}\label{eq12_3_9}
V_{\boldsymbol{m},\boldsymbol{k}_{\perp}} =&(2\pi)^{\frac{D}{2}}(-1)^{l-m_2}
\frac{i^{-l}}{\mathcal{C}_l^{\text{reg}}} C_{\boldsymbol{m}} \boldsymbol{Y}_{\boldsymbol{m}}( \boldsymbol{\theta}_k)^*e^{i\sqrt{k^2-k_{\perp}^2}L},
\end{align}where $\boldsymbol{\theta}_k=(\theta_1^k,\ldots,\theta_{D-1}^k)$ is defined by \eqref{eq12_3_8} with $k_1=\sqrt{k^2-k_{\perp}^2}$ and $k_{\perp}=k\sin\theta_1^k$. Here we have also used the fact that $C_n^{\nu}(-z)=(-1)^n C_n^{\nu}(z)$.

For the derivation of $W_{\boldsymbol{k}_{\perp},\boldsymbol{m}}$, first  notice that
\begin{align*}
j_0(kr)=a_D\int_{S_k^{D-1}} d\Omega_{D-1}^ke^{i\boldsymbol{k}\cdot\boldsymbol{r}},
\end{align*}
where
$$a_D=\frac{1}{(2\pi)^{\frac{D}{2}}}.$$
A counterpart for the outgoing wave is
\begin{align*}
h_0^{(1)}(kr)=2a_D\int_{\mathbb{R}^{D-1}}d\boldsymbol{k}_{\perp}\frac{e^{i\boldsymbol{k}\cdot\boldsymbol{r}}}{ k^{D-2}\sqrt{k^2-k_{\perp}^2}},
\end{align*}with $k_1=\sqrt{k^2-k_{\perp}^2}$.
Now
we will use the method in \cite{31,7}.
Using the fact that the normalized hyperspherical harmonics $C_{\boldsymbol{m}}\boldsymbol{Y}_{\boldsymbol{m}}(\boldsymbol{\theta})$ can be written as
\begin{align*}
C_{\boldsymbol{m}}\boldsymbol{Y}_{\boldsymbol{m}}(\boldsymbol{\theta})=H_{\boldsymbol{m}}\left(\frac{x_1}{r},\frac{x_2}{r},\ldots,\frac{x_D}{r}\right)
\end{align*}for some homogeneous polynomial $H_{\boldsymbol{m}}\left(x_1,\ldots,x_D\right)$ of degree $m_1=l$, we can define an operator $$\mathcal{H}_{\boldsymbol{m}}(\boldsymbol{\pa})=H_{\boldsymbol{m}}\left(\frac{\pa_{x_1}}{ik},\ldots,\frac{\pa_{x_D}}{ik}\right)$$which generalizes the operator $\mathcal{P}_{lm}$ defined in \cite{31}. It follows from definition that
\begin{align*}
\mathcal{H}_{\boldsymbol{m}}(\boldsymbol{\pa})e^{i\boldsymbol{k}\cdot\boldsymbol{r}}
=C_{\boldsymbol{m}}\boldsymbol{Y}_{\boldsymbol{m}}(\boldsymbol{\theta}_k)e^{i\boldsymbol{k}\cdot\boldsymbol{r}}.
\end{align*}
Hence, \eqref{eq12_3_4} can be written as
\begin{align*}
\varphi_{\boldsymbol{m}}^{\text{reg}}(\boldsymbol{x},k)=&\mathcal{C}_l^{\text{reg}}C_{\boldsymbol{m}}a_Di^l\int_{S^{D-1}_k}d\Omega_{D-1}^k\boldsymbol{Y}_{\boldsymbol{m}}(\boldsymbol{\theta}_k)
e^{i\boldsymbol{k}\cdot\boldsymbol{r}}\\
=&\mathcal{C}_l^{\text{reg}}a_Di^l\mathcal{H}_{\boldsymbol{m}}(\boldsymbol{\pa})\int_{S^{D-1}_k} d\Omega_{D-1}^ke^{i\boldsymbol{k}\cdot\boldsymbol{r}}\\
=&i^l\mathcal{C}_l^{\text{reg}}\mathcal{H}_{\boldsymbol{m}}(\boldsymbol{\pa})j_0(kr),
\end{align*} which says that $\varphi_{\boldsymbol{m}}^{\text{reg}}(\boldsymbol{x},k)$ can be obtained by applying the operator $\mathcal{H}_{\boldsymbol{m}}(\boldsymbol{\pa})$ on $j_0(kr)$. Since $j_{\nu}(z)$ and $h_{\nu}^{(1)}(z)$ satisfies the same differential equation, it follows that
\begin{align*}
\varphi_{\boldsymbol{m}}^{\text{out}}(\boldsymbol{x},k)=&i^l\mathcal{C}_l^{\text{out}}\mathcal{H}_{\boldsymbol{m}}(\boldsymbol{\pa})h_0^{(1)}(kr).
\end{align*}
Namely,
\begin{align*}
\varphi_{\boldsymbol{m}}^{\text{out}}(\boldsymbol{x},k)=&2i^l \mathcal{C}_l^{\text{out}} a_DC_{\boldsymbol{m}}\int_{\mathbb{R}^{D-1}}d\boldsymbol{k}_{\perp}\boldsymbol{Y}_{\boldsymbol{m}}(\boldsymbol{\theta}_k)\frac{e^{i\boldsymbol{k}\cdot\boldsymbol{r}}}{ k^{D-2}\sqrt{k^2-k_{\perp}^2}}.
\end{align*}
Compare to \eqref{eq12_3_7}, we find immediately that
\begin{align}\label{eq12_3_10}
W_{\boldsymbol{m},\boldsymbol{k}_{\perp}}(\boldsymbol{L})=&
\frac{2i^l(2\pi)^{D-1}a_D}{H^{D-1}}\mathcal{C}_l^{\text{out}}\frac{1}{k^{D-2}\sqrt{k^2-k_{\perp}^2}}
C_{\boldsymbol{m}}\boldsymbol{Y}_{\boldsymbol{m}}(\boldsymbol{\theta}_k)e^{i\sqrt{k^2-k_{\perp}^2}L}.
\end{align}

Substituting the expressions for $V_{\boldsymbol{m},\boldsymbol{k}_{\perp}}$ \eqref{eq12_3_9} and $W_{\boldsymbol{k}_{\perp},\boldsymbol{m}}$ \eqref{eq12_3_10} into the TGTG formula \eqref{eq12_3_1}, we find that the Casimir
 interaction energy between the sphere and the plate is given by
\begin{align}\label{eq12_3_13}
E_{\text{Cas}}=\frac{\hbar c}{2\pi}\int_0^{\infty} d \kappa \text{Tr}\,\ln\left(1-\mathbb{M}(\kappa)\right),
\end{align}
where the  $(\boldsymbol{m},\boldsymbol{m}')$  element of $\mathbb{M}$ is given by
\begin{align*}
M_{\boldsymbol{m},\boldsymbol{m}'}=&T_{1,\boldsymbol{m}}\frac{H^{D-1}}{(2\pi)^{D-1}}\int_{\mathbb{R}^{D-1}}d\boldsymbol{k}_{\perp} V_{\boldsymbol{m},\boldsymbol{k}_{\perp}}
T_{2,\boldsymbol{k}_{\perp}}W_{\boldsymbol{m}',\boldsymbol{k}_{\perp}}\\
=&\pi(-1)^{l+l'+m_2}i^{-m_2-m_2'}C_{\boldsymbol{m}}C_{\boldsymbol{m}'}T_{1,\boldsymbol{m}}\int_{0}^{\infty}dk_{\perp}\,k_{\perp}^{D-2}\int_0^{\pi}
d\theta_2^k\sin^{D-3}\theta_2^k\ldots\int_0^{\pi}d\theta_{D-2}^k \sin\theta_{D-2}^k\int_{-\pi}^{\pi}d\theta_{D-1}^k\\&
T_{2,\boldsymbol{k}_{\perp}}\frac{e^{-2L\sqrt{\kappa^2+k_{\perp}^2}}}{\kappa^{D-2}\sqrt{\kappa^2+k_{\perp}^2}}\left(\frac{k_{\perp}}{\kappa}\right)^{m_2+m_2'}
C_{l-m_2}^{m_{2}+\frac{D-2}{2}}\left(\frac{\sqrt{\kappa^2+k_{\perp}^2}}{\kappa}\right)C_{l'-m_2'}^{m_{2}'+\frac{D-2}{2}}\left(\frac{\sqrt{\kappa^2+k_{\perp}^2}}{\kappa}\right)e^{-im_{D-1}\theta_{D-1}}e^{im_{D-1}'\theta_{D-1}}
\\&\times\prod_{j=2}^{D-2}\sin^{m_{j+1}}\theta_j C_{m_j-m_{j+1}}^{m_{j+1}+\frac{D-j-1}{2}}(\cos\theta_j)\prod_{j'=2}^{D-2}\sin^{m_{j'+1}'}\theta_j C_{m_{j'}'-m_{j'+1}'}^{m_{j'+1}'+\frac{D-j'-1}{2}}(\cos\theta_j)\\
=& (-1)^{l+l'}2^{2m_2+D-3}  \delta_{\boldsymbol{m}_{\perp},\boldsymbol{m}_{\perp}'}\Gamma\left(m_2+\frac{D-2}{2}\right)^2
\sqrt{\frac{\left(l+\frac{D-2}{2}\right)\left(l'+\frac{D-2}{2}\right)(l-m_2)!(l'-m_2)!}{(l+m_2+D-3)!(l'+m_2+D-3)!}}
T_{1,\boldsymbol{m}}\\&\int_{0}^{\infty}dk_{\perp}\,
T_{2,\boldsymbol{k}_{\perp}}\frac{e^{-2L\sqrt{\kappa^2+k_{\perp}^2}}}{ \sqrt{\kappa^2+k_{\perp}^2}}\left(\frac{k_{\perp}}{\kappa}\right)^{2m_2 +D-2}
C_{l-m_2}^{m_{2}+\frac{D-2}{2}}\left(\frac{\sqrt{\kappa^2+k_{\perp}^2}}{\kappa}\right)C_{l'-m_2}^{m_{2}+\frac{D-2}{2}}\left(\frac{\sqrt{\kappa^2+k_{\perp}^2}}{\kappa}\right).
\end{align*}
Here $\boldsymbol{m}_{\perp}=(m_2,\ldots,m_{D-1})$. In the last row, we have used the orthogonality relation \eqref{eq12_3_11} to integrate out $\theta_2^k,\ldots, \theta_{D-1}^k$.
By the change of variable
\begin{align*}
k_{\perp}=\kappa\sinh\theta,
\end{align*}
we have
\begin{equation}\label{eq12_3_15}\begin{split}
M_{\boldsymbol{m},\boldsymbol{m}'}=&(-1)^{l+l'}  2^{2m_2+D-3}\delta_{\boldsymbol{m}_{\perp},\boldsymbol{m}_{\perp}'}\Gamma\left(m_2+\frac{D-2}{2}\right)^2
\sqrt{\frac{\left(l+\frac{D-2}{2}\right)\left(l'+\frac{D-2}{2}\right)(l-m_2)!(l'-m_2)!}{(l+m_2+D-3)!(l'+m_2+D-3)!}}
T_{1,l}T_{2 }\\&\int_{0}^{\infty}d\theta
 e^{-2\kappa L\cosh\theta}\left(\sinh\theta\right)^{2m_2 +D-2}
C_{l-m_2}^{m_{2}+\frac{D-2}{2}}\left(\cosh\theta\right)C_{l'-m_2}^{m_{2}+\frac{D-2}{2}}\left(\cosh\theta\right).
\end{split}\end{equation}
We take out the $T_2$ term from the integral since it is equal to $\pm 1$ for Dirichlet or Neumann boundary conditions. If one substitutes  $D=3$ in this expression and use the relation between Gegenbauer polynomials and associated Legendre functions \cite{29,30}, one recovers the expression derived in \cite{7} for $D=3$. The only difference is that when $D=3$, $m_2$ can take negative values. In that case, one has to replace the $m_2$ in the formula above by its absolute value.

Since $\mathbb{M}$ is diagonal in $\boldsymbol{m}_{\perp}=(m_2,\ldots,m_{D-1})$, we can simplify the trace in \eqref{eq12_3_13} as follows. When $D\geq 5$,
\begin{align}\label{eq12_4_1}
 \sum_{m_3=0}^{m_2}\ldots\sum_{m_{D-2}=0}^{m_{D-3}}\sum_{m_{D-1}=-m_{D-2}}^{m_{D-2}}1= \frac{(2m_2+D-3)(m_2+D-4)!}{(D-3)!m_2!}.
\end{align}When $D=4$,
\begin{align*}
\sum_{m_3=-m_2}^{m_2}1=2m_2+1
\end{align*}
which is equal to the right hand side of \eqref{eq12_4_1} when $D=4$.
Hence, when $D\geq 4$, the Casimir interaction energy \eqref{eq12_3_13} can be rewritten as
\begin{align}\label{eq12_3_14}
E_{\text{Cas}}=\frac{\hbar c}{2\pi}\int_0^{\infty} d \kappa \sum_{m=0}^{\infty}\frac{(2m+D-3)(m+D-4)!}{(D-3)!m!}\text{Tr}\,\ln\left(1-\mathbb{M}_{m}(\kappa)\right),
\end{align}
where the elements $M_{m;l,l'}$ of $\mathbb{M}_{m}$ is obtained from \eqref{eq12_3_15} by removing the factor $\delta_{\boldsymbol{m},\boldsymbol{m}'}$ and replacing $m_2$ with $m$. For fixed $m$, $l,l'$ ranges from $m$ to $\infty$.

Since the right hand side of \eqref{eq12_4_1} is equal to 2 when $D=3$, one can formally replace the summation   $\sum_{m=0}^{\infty}$ in \eqref{eq12_3_14} by $\sum_{m=0}^{\infty}\,'$ to obtain the $D=3$ case. Here the prime on the summation means that the term with $m=0$ is summed with weight $1/2$.

Finally, we would like to make a remark regarding the integral over $\theta$ in \eqref{eq12_3_15}. Recall that $C_n^{\nu}(z)$ is a polynomial of degree $n$ in $z$. Hence, if $l+l'$ is even,
\begin{equation}\label{eq12_4_2}
\left(\sinh\theta\right)^{2m_2 +D-2}
C_{l-m_2}^{m_{2}+\frac{D-2}{2}}\left(\cosh\theta\right)C_{l'-m_2}^{m_{2}+\frac{D-2}{2}}\left(\cosh\theta\right)\end{equation} can be written as a polynomial in $\sinh\theta$. If $l+l'$ is odd, \eqref{eq12_4_2} can be written as $\cosh\theta$ times a polynomial in $\sinh\theta$.
Then the integral over $\theta$ can be explicitly computed using \cite{32}:
\begin{equation}\label{eq12_4_3}
\begin{split}
\int_0^{\infty} d\theta e^{-2\kappa L\cosh\theta}\sinh^{n}\theta=&\frac{(\kappa L)^{-\frac{n}{2}}}{\sqrt{\pi}}\Gamma\left(\frac{n+1}{2}\right)K_{\frac{n}{2}}\left(2\kappa L\right),\\
\int_0^{\infty} d\theta e^{-2\kappa L\cosh\theta}\sinh^{n}\theta\cosh\theta=&\frac{(\kappa L)^{-\frac{n}{2}}}{\sqrt{\pi}}\Gamma\left(\frac{n+1}{2}\right)K_{\frac{n+2}{2}}\left(2\kappa L\right).
\end{split}
\end{equation}Hence, the integral over $\theta$ in \eqref{eq12_3_15} can be written as a linear combination of Bessel functions $K_{\nu}(2\kappa L)$. This might be more convenient for numerical computations.

\section{Large separation asymptotic behavior}
In this section, we derive the asymptotic behavior of the Casimir interaction energy \eqref{eq12_3_14} when the separation between the sphere and the plate is large, i,e, when $L\gg a$.

Expanding the logarithm and the trace in \eqref{eq12_3_14}, we have
\begin{align}\label{eq12_3_16}
E_{\text{Cas}}=-\frac{\hbar c}{2\pi}\sum_{s=0}^{\infty}\frac{1}{s+1}\int_0^{\infty} d \kappa \sum_{m=0}^{\infty}\frac{(2m+D-3)(m+D-4)!}{(D-3)!m!}
\sum_{l_0=m}^{\infty}\sum_{l_1=m}^{\infty}\ldots\sum_{l_s=m}^{\infty} \prod_{j=0}^sM_{m;l_j,l_{j+1}},
\end{align}with the convention that $l_{s+1}=l_0$.

Making a change of variables $\kappa\mapsto \kappa/L$, one can see that the leading asymptotic behavior of the Casimir interaction energy is governed by the small $z=\kappa a/L$ asymptotic behaviors of   \begin{equation*}
\begin{split}
&T_{1,l}^{\text{D}} =\frac{I_{l+\frac{D-2}{2}}(z)}{K_{l+\frac{D-2}{2}}(z)},\\
&T_{1,l}^{\text{N}} =\frac{-\frac{D-2}{2}I_{l+\frac{D-2}{2}}(z)+zI_{l+\frac{D-2}{2}}'(z)}
{-\frac{D-2}{2}K_{l+\frac{D-2}{2}}(z)+z K_{l+\frac{D-2}{2}}'(z)}.\end{split}
\end{equation*}

As $z\rightarrow 0$, we have the following asymptotic behaviors:
\begin{align*}
I_{\nu}(z)\sim & \frac{1}{\Gamma(\nu+1)}\left(\frac{z}{2}\right)^{\nu},\\
K_{\nu}(z)\sim &\frac{\Gamma(\nu)}{2}\left(\frac{z}{2}\right)^{-\nu}.
\end{align*}Hence,
\begin{equation}\label{eq12_4_7}\begin{split}
T_{1,l}^{\text{D}}\sim & \frac{1}{2^{2l+D-3}\Gamma\left(l+\frac{D-2}{2}\right)\Gamma\left(l+\frac{D}{2}\right)}z^{2l+D-2},\\
T_{1,l}^{\text{N}}\sim & -\frac{l}{l+D-2}\frac{1}{2^{2l+D-3}\Gamma\left(l+\frac{D-2}{2}\right)\Gamma\left(l+\frac{D}{2}\right)}z^{2l+D-2},\quad\text{if}\;l\neq 0\\
T_{1,0}^{\text{N}}\sim &-\frac{1}{(D-2)2^{D-2}\Gamma\left(\frac{D-2}{2}\right)\Gamma\left(\frac{D+2}{2}\right)}z^{D}.
\end{split}\end{equation}
These show that when $L\gg a$, the leading terms of the Casimir interaction energy come  from the terms with small $l$. When the sphere is imposed with Dirichlet boundary condition, the dominating term is the term with $l=0$. When the sphere is imposed with Neumann boundary condition, both the terms with $l=0$ and $l=1$ contribute a term with the same order in $L$.

Now we compute the leading term for the Casimir interaction energy for $L\gg a$. If the sphere is imposed with Dirichlet boundary conditions, the dominating term is the term with $s=0$ and $m=l_0=0$ in \eqref{eq12_3_16}. Namely,
\begin{align*}
 E_{\text{Cas}}^{\text{DD/DN}}\sim -\frac{\hbar c}{2\pi}\int_0^{\infty} d \kappa M_{0;0,0}^{\text{DD/DN}}(\kappa).
\end{align*}Now, using $C_0^{\nu}(z)=1$ and \eqref{eq12_4_3} and \eqref{eq12_4_7}, we have
\begin{align*}
M_{0;0,0}^{\text{DD/DN}}\sim &T_2^{\text{D/N}}\frac{a^{D-2}}{\sqrt{\pi}(D-3)!L^{\frac{D-2}{2}}}\Gamma\left(\frac{D-1}{2}\right)\kappa^{\frac{D-2}{2}}K_{\frac{D-2}{2}}(2\kappa L).
\end{align*}
Integrating over $\kappa$ gives
\begin{align}\label{eq12_4_5}
 E_{\text{Cas}}^{\text{DD/DN}}
 \sim &\mp \frac{\hbar c a^{D-2}}{ 8\pi (D-3)!L^{D-1}} \Gamma\left(\frac{D-1}{2}\right)^2.
\end{align}

If the sphere is imposed with Neumann boundary conditions, the dominating term is the term with $s=0$ and $m=l_0=0$ or $l_0=1$ and $m=0$ or $1$ in \eqref{eq12_3_16}. Namely,
\begin{align*}
 E_{\text{Cas}}^{\text{NN/ND}}\sim -\frac{\hbar c}{2\pi}\int_0^{\infty} d \kappa \left(M_{0;0,0}^{\text{NN/ND}}(\kappa)+M_{0;1,1}^{\text{NN/ND}}(\kappa)+(D-1)M_{1;1,1}^{\text{NN/ND}}(\kappa)\right).
\end{align*}Now,
\begin{align*}
M_{0;0,0}^{\text{NN/ND}}\sim &-T_2^{\text{N/D}} \frac{a^{D}}{\sqrt{\pi}D(D-2)!L^{\frac{D-2}{2}}}\Gamma\left(\frac{D-1}{2}\right)\kappa^{\frac{D+2}{2}}K_{\frac{D-2}{2}}(2\kappa L),
\end{align*}
\begin{align*}
M_{1;1,1}^{\text{NN/ND}}\sim &-T_2^{\text{N/D}} \frac{a^{D}}{\sqrt{\pi}(D-1)(D-1)!L^{\frac{D}{2}}}\Gamma\left(\frac{D+1}{2}\right)\kappa^{\frac{D}{2}}K_{\frac{D}{2}}(2\kappa L),
\end{align*}and using $C_1^{\nu}(z)=2\nu z$,
\begin{align*}
M^{\text{NN/ND}}_{0;1,1}\sim &-T_2^{\text{N/D}} \frac{a^{D}}{\sqrt{\pi} (D-1)!L^{\frac{D}{2}}}\left(\Gamma\left(\frac{D+1}{2}\right)\kappa^{\frac{D}{2}}K_{\frac{D}{2}}(2\kappa L)
 +\Gamma\left(\frac{D-1}{2}\right)L\kappa^{\frac{D+2}{2}}K_{\frac{D-2}{2}}(2\kappa L)
 \right).
\end{align*}
Integrating over $\kappa$ gives \begin{align}\label{eq12_4_6}
 E_{\text{Cas}}^{\text{NN/ND}}\sim & \mp\frac{\hbar ca^D}{16\pi D!L^{D+1}}\Gamma\left(\frac{D-1}{2}\right)
\Gamma\left(\frac{D+1}{2}\right)(2D^2-1).
\end{align}

Hence, we find that when $L\gg a$,  the leading term of the Casimir interaction energy is of order $L^{-(D-1)}$ for Dirichlet boundary condition, and of order $L^{-(D+1)}$ for Neumann boundary conditions. When $D=3$, \eqref{eq12_4_5} and \eqref{eq12_4_6} give respectively
\begin{equation*}
\begin{split}
 E_{\text{Cas}}^{\text{DD/DN}}
 \sim &\mp \frac{\hbar c a }{ 8\pi L^{2}},\\
   E_{\text{Cas}}^{\text{NN/ND}}\sim & \mp\frac{17\hbar ca^3}{96\pi L^{4}},
\end{split}
\end{equation*}which agree with the results derived in \cite{33}.

In general, we can write
\begin{align*}
E_{\text{Cas}}^{\text{DD/DN}}\sim &\alpha_D^{\text{DD/DN}}\frac{\hbar c a^{D-2}}{ L^{D-1}}\\
E_{\text{Cas}}^{\text{NN/ND}}\sim &\alpha_D^{\text{ND/NN}}\frac{\hbar c a^{D}}{L^{D+1}}.
\end{align*}$\alpha_D$ are pure numbers that depend on $D$. Obviously, $\alpha_D^{\text{DN}}=-\alpha_D^{\text{DD}}$ and $\alpha_D^{\text{ND}}=-\alpha_D^{\text{NN}}$. The values of $\alpha_D$ for $3\leq D\leq 12$ are tabulated in Table \ref{t1} in Appendix \ref{a1}.

\section{Proximity force approximation}
In this section, we compute the proximity force approximation (PFA) to the Casimir interaction energy. In the next section, we will compute the small separation asymptotic behavior of the Casimir interaction energy from \eqref{eq12_3_14} and compare to the result obtained in this section.

In $(D+1)$-dimensional Minkowski spacetime, the Casimir   energy density between two parallel plates  both subject to Dirichlet or Neumann boundary conditions is given by \cite{9}:
\begin{align*}
\mathcal{E}_{\text{Cas}}^{\parallel,\text{DD/NN}} (d)=-\hbar c\frac{\Gamma\left(\frac{D+1}{2}\right)\zeta(D+1)}{2^{D+1}\pi^{\frac{D+1}{2}}}\frac{1}{d^{D}}=\frac{b_D^{\text{DD/NN}}}{d^{D}},
\end{align*}where $d$ is the distance between the two plates, and $\zeta(z)$ is the Riemann zeta function. When one plate is imposed with Dirichlet boundary conditions, and one is imposed with Neumann boundary conditions, the Casimir energy density is
\begin{align*}
\mathcal{E}_{\text{Cas}}^{\parallel,\text{DN/ND}} (d)=\hbar c\left(1-2^{-D}\right)\frac{\Gamma\left(\frac{D+1}{2}\right)\zeta(D+1)}{2^{D+1}\pi^{\frac{D+1}{2}}}\frac{1}{d^{D}}=\frac{b_D^{\text{DN/ND}}}{d^{D}},
\end{align*}

The distance between the point  $(a,\theta_1,\ldots,\theta_{D-1})$ (in spherical coordinates) on the sphere to the plate $x_1=L$ is
$$d(\boldsymbol{\theta})=d(\theta_1)=L-a\cos\theta_1=d+a(1-\cos\theta_1),$$where $d=L-a$ is the shortest distance between the sphere and the plate.

Hence, the proximity force approximation to the Casimir interaction energy between the sphere and the plate is
\begin{align*}
E_{\text{Cas}}^{\text{PFA}}=& a^{D-1}\int_{S^{D-1}}\mathcal{E}_{\text{Cas}}^{\parallel} (d(\boldsymbol{\theta}))d\Omega_{D-1}\\
=&a^{D-1}b_D\frac{2\pi^{\frac{D-1}{2}}}{\Gamma\left(\frac{D-1}{2}\right)}\int_0^{\pi}\frac{d\theta_1\sin^{D-2}\theta_1}{\left(d+a(1-\cos\theta_1)\right)^{D}}.
\end{align*}
We want to compute the leading term of the proximity force approximation when $$\vep=\frac{d}{a}\ll 1.$$
Making a change of variable
\begin{align*}
v=\frac{d+a(1-\cos\theta_1)}{d},
\end{align*}
we have
\begin{align*}
E_{\text{Cas}}^{\text{PFA}}\sim & b_D\frac{2\pi^{\frac{D-1}{2}}}{\Gamma\left(\frac{D-1}{2}\right)}\frac{a}{ d^{\frac{D+1}{2}}}\int_1^{\frac{2a+d}{d}} dv
\frac{\left(2a+d-dv\right)^{\frac{D-3}{2}}(v-1)^{\frac{D-3}{2}}}{v^D}\\
\sim & b_D\frac{(2\pi)^{\frac{D-1}{2}}}{\Gamma\left(\frac{D-1}{2}\right)}\frac{a^{\frac{D-1}{2}}}{d^{\frac{D+1}{2}}}\int_1^{\infty} dv
\frac{ (v-1)^{\frac{D-3}{2}}}{v^D}\\
= & b_D\frac{ \pi^{\frac{D}{2}}}{2^{\frac{D-1}{2}}\Gamma\left(\frac{D}{2}\right)}\frac{1}{a\vep^{\frac{D+1}{2}}}.
\end{align*}In the last line, we have used the formula
\begin{equation}\label{eq12_5_5}
\Gamma(2z)=\frac{2^{2z-1}}{\sqrt{\pi}}\Gamma(z)\Gamma\left(z+\frac{1}{2}\right).
\end{equation}
Hence, we find that the proximity force approximation to the Casimir interaction energy between a sphere and a plate is
\begin{equation}\label{eq12_5_10}\begin{split}
E_{\text{Cas}}^{\text{PFA, DD/NN}}\sim &-\frac{\Gamma\left(\frac{D+1}{2}\right)\zeta(D+1)}{2^{\frac{3D+1}{2}}\sqrt{\pi}\Gamma\left(\frac{D}{2}\right)} \frac{\hbar c}{a\vep^{\frac{D+1}{2}}},\\
E_{\text{Cas}}^{\text{PFA, DN/ND}}\sim &(1-2^{-D})\frac{\Gamma\left(\frac{D+1}{2}\right)\zeta(D+1)}{2^{\frac{3D+1}{2}}\sqrt{\pi}\Gamma\left(\frac{D}{2}\right)} \frac{\hbar c}{a\vep^{\frac{D+1}{2}}}.
\end{split}\end{equation}
 \section{Analytic computation of leading and next-to-leading order term of small separation asymptotic behavior}
In this section, we compute the small separation asymptotic behavior of the Casimir interaction energy from the formula \eqref{eq12_3_14}. The method we use is the perturbation method applied in \cite{3} for the cylinder-plate configuration, and later used in \cite{19,20,21,24} for the sphere-plate configuration.

The Casimir interaction energy is given by \eqref{eq12_3_16} with the matrix element $M_{m;l,l'}$ given by \eqref{eq12_3_15}.

In \cite{19,20,21}, the  integration over $\theta$ in \eqref{eq12_3_15} has been performed and the $M_{m;l,l'}$ is expressed in terms of $3j$-symbols. Hence, to find the asymptotic behavior of the Casimir interaction energy, an integral formula of the $3j$-symbol is used to find its asymptotic behavior. In \cite{24}, we could not perform the integration over $\theta$ in $M_{m;l,l'}$, but we showed that by using an integral formula for the associated Legendre function, we could find the asymptotic behavior of the integral which  is not any complicated than finding the asymptotic behavior of the $3j$-symbol in \cite{19,20,21}.

In our present case, we have to find an analogous integral formula for the Gegenbauer polynomials. In fact, using the Rodrigue's formula \cite{29,32}:
\begin{align*}
C_n^{\nu}(z)=&\frac{\Gamma\left(\nu+\frac{1}{2}\right)\Gamma(n+2\nu)}{2^nn!\Gamma(2\nu)\Gamma\left(\nu+\frac{1}{2}+n\right)}
(z^2-1)^{\frac{1}{2}-\nu}\frac{d^n}{dz^n}\left(z^2-1\right)^{n+\nu-\frac{1}{2}},
\end{align*}one can use the same method as for associated Legendre function (see page 303 in \cite{34}) to show that
\begin{equation}\label{eq12_5_2}\begin{split}
(z^2-1)^{\frac{\nu}{2}-\frac{1}{4}}C_n^{\nu}(z)
=& \frac{\Gamma\left(\nu+\frac{1}{2}\right)\Gamma(n+2\nu)}{\Gamma(2\nu)\Gamma\left(\nu+\frac{1}{2}+n\right)}
 \frac{2^{\nu-\frac{3}{2}}}{  \pi  }\int_{-\pi}^{\pi} \left(z+\sqrt{z^2-1}\cos\varphi\right)^{n+\nu-\frac{1}{2}}e^{i\left(\nu-\frac{1}{2}\right)\varphi}d\varphi.
\end{split}\end{equation}

Let $$\tilde{m}=m+\frac{D-3}{2}.$$When $m$ ranges over nonnegative integers,  $\tilde{m}$ ranges over $\displaystyle \frac{D-3}{2}, \frac{D-1}{2}, \frac{D+1}{2}, \ldots$, which are integers if and only if $D$ is odd. The Casimir interaction energy \eqref{eq12_3_16} can be rewritten as
\begin{align}\label{eq12_5_1}
E_{\text{Cas}}=-\frac{\hbar c}{(D-3)! \pi }\sum_{s=0}^{\infty}\frac{1}{s+1}\int_0^{\infty} d \kappa \sum_{\tilde{m}}\frac{\tilde{m}\left(\tilde{m}+\frac{D-5}{2}\right)!}{\left(\tilde{m}-\frac{D-3}{2}\right)!}
\sum_{l=\tilde{m}-\frac{D-3}{2}}^{\infty}\sum_{l_1=\tilde{m}-\frac{D-3}{2}-l}^{\infty}\ldots\sum_{l_s=\tilde{m}-\frac{D-3}{2}-l}^{\infty} \prod_{j=0}^sM_{\tilde{m};l_j,l_{j+1}},
\end{align}where we have replaced $l_0$ with $l$ and  $l_j$ with $l+l_j$ for $1\leq j\leq s$. $M_{\tilde{m};l_j,l_{j+1}}$ is then given by
\begin{equation}\label{eq12_5_4}\begin{split}
M_{\tilde{m}; l_j,l_{j+1}}=&(-1)^{l_j+l_{j+1}}  2^{2\tilde{m}} \Gamma\left(\tilde{m}+\frac{1}{2}\right)^2
\sqrt{\frac{\left(l+l_j+\frac{D-2}{2}\right)\left(l+l_{j+1}+\frac{D-2}{2}\right)\Gamma\left(l+l_j+\frac{D-1}{2}-\tilde{m}\right)\Gamma\left(l+l_{j+1}+\frac{D-1}{2}-\tilde{m}\right)}
{\Gamma\left(l+l_j+\frac{D-1}{2}+\tilde{m}\right)\Gamma\left(l+l_j+\frac{D-1}{2}+\tilde{m}\right)}}
\\&\times T_{1,l+l_j}T_{2 }\int_{0}^{\infty}d\theta
 e^{-2\kappa L\cosh\theta}\sinh\theta\left(\sinh\theta\right)^{2\tilde{m}}
C_{l+l_j+\frac{D-3}{2}-\tilde{m}}^{\tilde{m}+\frac{1}{2}}\left(\cosh\theta\right)C_{l+l_{j+1}+\frac{D-3}{2}-\tilde{m}}^{\tilde{m}+\frac{1}{2}}\left(\cosh\theta\right).
\end{split}\end{equation}

Using the formula \eqref{eq12_5_5},   eq. \eqref{eq12_5_2} shows that
\begin{align*}
\sinh^{\tilde{m}}\theta C_{l+l_j+\frac{D-3}{2}-\tilde{m}}^{\tilde{m}+\frac{1}{2}}(\cosh\theta)=&\frac{\Gamma\left(l+l_j+\frac{D-1}{2}+\tilde{m}\right)}{\Gamma\left(\tilde{m}+\frac{1}{2}\right)
\Gamma\left(l+l_j+\frac{D-1}{2}\right)}
 \frac{1}{ 2^{\tilde{m}}\sqrt{\pi}  }\int_{-\frac{\pi}{2}}^{\frac{\pi}{2}} \left(\cosh\theta+\sinh\theta\cos2\varphi\right)^{l+l_j+\frac{D-3}{2}}e^{2i\tilde{m}\varphi}d\varphi.
\end{align*}
Using binomial expansion,
\begin{align*}
\left(\cosh\theta+\sinh\theta\cos2\varphi\right)^{l+l_j+\frac{D-3}{2}}=&\sum_{k=0}^{\infty} \frac{1}{k!}\frac{\Gamma\left(l+l_j+\frac{D-1}{2}\right)}{\Gamma\left(l+l_j+\frac{D-1}{2}-k\right)}\exp\left(\left(l+l_j+\frac{D-3}{2}-2k\right)\theta\right)\left(\cos\varphi\right)^{2l+2l_j+D-3-2k}
\sin^{2k}\varphi.
\end{align*}
Therefore, eq. \eqref{eq12_5_4} can be rewritten as
\begin{equation}\label{eq12_5_6}\begin{split}
M_{\tilde{m}; l_j,l_{j+1}}=&\frac{(-1)^{l_j+l_{j+1}}}{\pi}
T_{1,l+l_j}T_{2 }\sum_{k=0}^{\infty}\sum_{k'=0}^{\infty}\frac{1}{k!}\frac{1}{k'!}\mathcal{N}_{\tilde{m};l_j,l_{j+1};k,k'} \int_{0}^{\infty}d\theta
 e^{-2\kappa L\cosh\theta}\sinh\theta e^{\left(2l+l_j+l_{j+1}+D-3-2k-2k'\right)\theta}\\
 &\times \int_{-\frac{\pi}{2}}^{\frac{\pi}{2}}d\varphi \left(\cos\varphi\right)^{2l+2l_j+D-3-2k}
\sin^{2k}\varphi e^{2i\tilde{m}\varphi}d\varphi \int_{-\frac{\pi}{2}}^{\frac{\pi}{2}}d\varphi' \left(\cos\varphi'\right)^{2l+2l_j+D-3-2k}
\sin^{2k}\varphi' e^{2i\tilde{m}\varphi'}d\varphi',
\end{split}\end{equation}
where
\begin{align*}
\mathcal{N}_{\tilde{m};l_j,l_{j+1};k,k'}=&\sqrt{\left(l+l_j+\frac{D-2}{2}\right)\left(l+l_{j+1}+\frac{D-2}{2}\right)}\\&\times \frac{\sqrt{ \Gamma\left(l+l_j+\frac{D-1}{2}-\tilde{m}\right)\Gamma\left(l+l_{j+1}+\frac{D-1}{2}-\tilde{m}\right)
\Gamma\left(l+l_j+\frac{D-1}{2}+\tilde{m}\right)\Gamma\left(l+l_j+\frac{D-1}{2}+\tilde{m}\right)}}{\Gamma\left(l+l_j+\frac{D-1}{2}-k\right)\Gamma\left(l+l_{j+1}+\frac{D-1}{2}-k'\right)}.
\end{align*}From these expressions, it is obvious that $M_{\tilde{m}; l_j,l_{j+1}}$ is invariant of we replace $\tilde{m}$ by $-\tilde{m}$.

Now we can find the asymptotic behavior of the Casimir interaction energy as in \cite{24}.
Let $\theta_0$ be defined so that
\begin{align*}
\sinh\theta_0=\frac{l}{\kappa a}.
\end{align*}Make a change of variable $\theta\mapsto \theta+\theta_0$ in \eqref{eq12_5_6}. Also, let
$$\kappa=\frac{l\sqrt{1-\tau^2}}{a\tau}.$$
When $\vep =d/a\ll 1$, the leading contribution to the Casimir interaction energy comes from terms with:
\begin{align*}
l\sim \frac{1}{\vep}, \quad l_j\sim\frac{1}{\sqrt{\vep}},\quad \tilde{m}\sim\frac{1}{\sqrt{\vep}},\quad \theta\sim \sqrt{\vep},\quad \tau\sim 1.
\end{align*}Hence, we are counting the order $l, l_j, \tilde{m}, \theta$ and $\tau$ as $1/\vep, 1/\sqrt{\vep}, 1/\sqrt{\vep}, \sqrt{\vep}$ and $1$ respectively. After performing the summation over $k, k'$ and the integration over $\varphi, \varphi'$ and $\theta$, we obtain an expression of the form
\begin{align}\label{eq12_5_8}
M_{\tilde{m};l_j,l_{j+1}}\sim &  \sigma\frac{C^{l_j-l_{j+1}}}{2}\sqrt{\frac{\tau}{\pi l}}\exp\left(-\frac{\tilde{m}^2}{l\tau} -\frac{2l\vep}{\tau}-\frac{\tau(l_j-l_{j+1})^2}{4l}\right)\left(1+\mathcal{A}_{j,1} +\mathcal{A}_{j,2}\right)\left(1+\mathcal{J} \right),
\end{align}
where $\mathcal{A}_{j,1}$ and $\mathcal{A}_{j,2}$ are respectively terms of order $\sqrt{\vep}$ and $\vep$, and $\mathcal{J}$ is a term of order $\vep$ which depends on the boundary conditions on the sphere:
\begin{equation}\label{eq12_6_1}\begin{split}
\mathcal{J}^{\text{D}}=&\frac{1}{l}\left(\frac{\tau}{4}-\frac{5\tau^3}{12}\right),\\
\mathcal{J}^{\text{N}}=&\frac{1}{l}\left(-\left(D-\frac{5}{4}\right)\tau+\frac{7\tau^3}{12}\right).
\end{split}\end{equation}
The factor $\sigma$ depends on the boundary conditions on both the sphere and the plate. For DD and NN boundary conditions, $\sigma=1$. For DN or ND boundary conditions, $\sigma=-1$.

Due to the exponential term in \eqref{eq12_5_8}, in \eqref{eq12_5_1}, one can replace the summations over $l, \tilde{m}$ and $l_j$ by integrations. Namely, for first two leading order terms,
\begin{align}\label{eq12_5_9}
E_{\text{Cas}}\sim -\frac{\hbar c}{(D-3)! \pi a}\sum_{s=0}^{\infty}\frac{1}{s+1}  \int_0^{\infty} dl \, l\int_0^{1}\frac{d\tau}{\tau^2\sqrt{1-\tau^2}} \int_0^{\infty}d\tilde{m}\frac{\tilde{m}\left(\tilde{m}+\frac{D-5}{2}\right)!}{\left(\tilde{m}-\frac{D-3}{2}\right)!}
 \int_{-\infty}^{\infty}dl_1\ldots\int_{-\infty}^{\infty} dl_s\prod_{j=0}^sM_{\tilde{m};l_j,l_{j+1}}.
\end{align}
A major difference with the $D=3$ case is the appearance of the term $$\frac{\tilde{m}\left(\tilde{m}+\frac{D-5}{2}\right)!}{\left(\tilde{m}-\frac{D-3}{2}\right)!}.$$ Obviously, this is a polynomial of degree $D-3$ in $\tilde{m}$, and straightforward computation gives
\begin{align*}
\frac{\tilde{m}\left(\tilde{m}+\frac{D-5}{2}\right)!}{\left(\tilde{m}-\frac{D-3}{2}\right)!}=  \tilde{m}^{D-3}-\frac{(D-3)(D-4)(D-5)}{24}\tilde{m}^{D-5}+\ldots.
\end{align*}We only need the first  two leading terms to find the leading order and next-to-leading order terms of the Casimir interaction energy.

Putting everything into \eqref{eq12_5_9} and integrating over $l_j$, $1\leq j\leq s$, we find that
\begin{align*}
E_{\text{Cas}}\sim &-\frac{\hbar c}{2(D-3)!\pi a}\sum_{s=0}^{\infty}\frac{\sigma^{s+1}}{(s+1)^{\frac{3}{2}}}\int_{0}^{\infty}dl \,l\int_0^{1} \frac{ d\tau}{\tau^2\sqrt{1-\tau^2}}\int_{0}^{\infty}d\tilde{m}  \left(\tilde{m}^{D-3}-\frac{(D-3)(D-4)(D-5)}{24}\tilde{m}^{D-5}\right)  \\
&\times\sqrt{\frac{\tau}{\pi l}}\exp\left(-\frac{\tilde{m}^2(s+1)}{l\tau} -\frac{2l(s+1)\vep}{\tau} \right)  \left(1  +\mathcal{B}\right)\left(1+(s+1)\mathcal{J}\right),
\end{align*}where $\mathcal{B}$ is a term of order $\vep$.

Hence, we find that the leading order term of the Casimir interaction energy is
\begin{align*}
E_{\text{Cas}}^0=&-\frac{\hbar c}{2(D-3)!\pi a}\sum_{s=0}^{\infty}\frac{\sigma^{s+1}}{(s+1)^{\frac{3}{2}}}\int_{0}^{\infty}dl\,l\int_0^{1} \frac{ d\tau}{\tau^2\sqrt{1-\tau^2}} \int_{0}^{\infty}d\tilde{m}   \tilde{m}^{D-3}   \sqrt{\frac{\tau}{ \pi l}}\exp\left(-\frac{\tilde{m}^2(s+1)}{l\tau} -\frac{2l(s+1)\vep}{\tau} \right). \end{align*}
After integrating over $\tilde{m}$, $l$ and $\tau$, we have
\begin{align*}
E_{\text{Cas}}^0
=&-\frac{\Gamma\left(\frac{D+1}{2}\right)  }{ 2^{\frac{3D+1}{2}}\sqrt{\pi}\Gamma\left(\frac{D}{2}\right)}\frac{\hbar c}{  a\vep^{\frac{D+1}{2}}} \sum_{s=0}^{\infty}\frac{\sigma^{s+1}}{(s+1)^{D+1}}.
\end{align*}
Thus, using the definition of Riemann zeta function $\zeta(s)=\sum_{n=1}^{\infty} 1/n^s$, we find that the leading order term of the Casimir interaction energy is
\begin{align*}
E_{\text{Cas}}^{0,\text{DD/NN}}
=&-\frac{\Gamma\left(\frac{D+1}{2}\right)\zeta(D+1)  }{ 2^{\frac{3D+1}{2}}\sqrt{\pi}\Gamma\left(\frac{D}{2}\right)}\frac{\hbar c}{  a\vep^{\frac{D+1}{2}}},\\
E_{\text{Cas}}^{0,\text{DN/ND}}
=&(1-2^{-D})\frac{\Gamma\left(\frac{D+1}{2}\right)\zeta(D+1)  }{ 2^{\frac{3D+1}{2}}\sqrt{\pi}\Gamma\left(\frac{D}{2}\right)}\frac{\hbar c}{  a\vep^{\frac{D+1}{2}}},
\end{align*}which agree with PFA \eqref{eq12_5_10}. We see that the leading order term is of order $\vep^{-(D+1)/2}$. Let us write
\begin{align*}
E_{\text{Cas}}^{0}=\beta_D\frac{\hbar c}{a\vep^{\frac{D+1}{2}}},
\end{align*}so that $\beta_D$ is a pure number. Obviously, $\beta_D^{\text{DD}}=\beta_D^{\text{NN}}$ and $\beta_D^{\text{DN}}=\beta_D^{\text{ND}}$. The values of $\beta_D$ for $3\leq D\leq 12$ is tabulated in Table \ref{t1} in Appendix \ref{a1}.

For the next to leading order term $E_{\text{Cas}}^1$, there are two contributions, one is
\begin{align*}
E_{\text{Cas}}^{1,1}=&\frac{(D-3)(D-4)(D-5)\hbar c}{48\pi (D-3)! a}\sum_{s=0}^{\infty}\frac{\sigma^{s+1}}{(s+1)^{\frac{3}{2}}}\int_{0}^{\infty}dl \,l \int_0^{1} \frac{d\tau}{\tau^2\sqrt{1-\tau^2}}\\&\times\int_{0}^{\infty}d\tilde{m}   \tilde{m}^{D-5}   \sqrt{\frac{\tau}{ \pi l}}\exp\left(-\frac{\tilde{m}^2(s+1)}{l\tau} -\frac{2l(s+1)\vep}{\tau} \right) \end{align*}
which vanishes for $D=3,4,5$, and the other is
\begin{align*}
E_{\text{Cas}}^{1,2}=&-\frac{\hbar c}{2\pi (D-3)! a}\sum_{s=0}^{\infty}\frac{\sigma^{s+1}}{(s+1)^{\frac{3}{2}}}\int_{0}^{\infty}dl\,l\int_0^{1} \frac{d\tau}{\tau^2\sqrt{1-\tau^2}} \int_{0}^{\infty}d\tilde{m}   \tilde{m}^{D-3}   \\&\times\sqrt{\frac{\tau}{ \pi l}}\exp\left(-\frac{\tilde{m}^2(s+1)}{l\tau} -\frac{2l(s+1)\vep}{\tau} \right)\left(\mathcal{B}+(s+1)\mathcal{J}\right). \end{align*}
Writing
\begin{align*}
E_{\text{Cas}}=E_{\text{Cas}}^0+E_{\text{Cas}}^1+\ldots=E_{\text{Cas}}^0\left(1+\vartheta_E\vep+o(\vep)\right),
\end{align*}so that
\begin{align*}
\vartheta_E=\frac{1}{\vep} \frac{E_{\text{Cas}}^{1}}{E_{\text{Cas}}^0}
\end{align*}
measures the first order deviation from the proximity force approximation. We find that $\vartheta_1$ is a pure number that depends on space dimension $D$ and the boundary conditions on the sphere and the plate. 
For $D=4$,
\begin{equation*}
\begin{split}
\vartheta_E^{\text{DD}}=&\frac{5}{12}-\frac{7}{18}\frac{\zeta(3)}{\zeta(5)},\\
\vartheta_E^{\text{DN}}=&\frac{5}{12}-\frac{14}{45}\frac{\zeta(3)}{\zeta(5)},\\
\vartheta_E^{\text{ND}}=&\frac{5}{12}-\frac{122}{45}\frac{\zeta(3)}{\zeta(5)},\\
\vartheta_E^{\text{NN}}=&\frac{5}{12}-\frac{61}{18}\frac{\zeta(3)}{\zeta(5)};
\end{split}
\end{equation*}while for $D=3$ or $D\geq 5$,
\begin{equation}\label{eq12_5_12}
\begin{split}
\vartheta_E^{\text{DD}}=&\frac{D+1}{12}-\frac{ (D-2)(D-3)}{3D }\frac{\zeta(D-1)}{\zeta(D+1)},\\
\vartheta_E^{\text{DN}}=&\frac{D+1}{12}-\frac{2^D-4}{2^D-1}\frac{ (D-2)(D-3)}{3D }\frac{\zeta(D-1)}{\zeta(D+1)},\\
\vartheta_E^{\text{ND}}=&\frac{D+1}{12}-\frac{2^D-4}{2^D-1} \frac{(D^2+7D-6)}{3D }  \frac{\zeta(D-1)}{\zeta(D+1)},\\
\vartheta_E^{\text{NN}}=&\frac{D+1}{12}- \frac{(D^2+7D-6)}{3D }  \frac{\zeta(D-1)}{\zeta(D+1)}.
\end{split}
\end{equation}When $D=3$, we recover the results in \cite{21,35}.

For the Casimir interaction force, it follows that
\begin{align*}
F_{\text{Cas}}=F_{\text{Cas}}^0+F_{\text{Cas}}^1+\ldots=F_{\text{Cas}}^0\left(1+\vartheta_F\vep+o(\vep)\right),
\end{align*}where
\begin{align*}
\vartheta_F=\frac{D-1}{D+1}\vartheta_E.
\end{align*}In particular, $\vartheta_F$ has the same sign as $\vartheta_E$.

\begin{figure}[h]
\epsfxsize=0.6\linewidth \epsffile{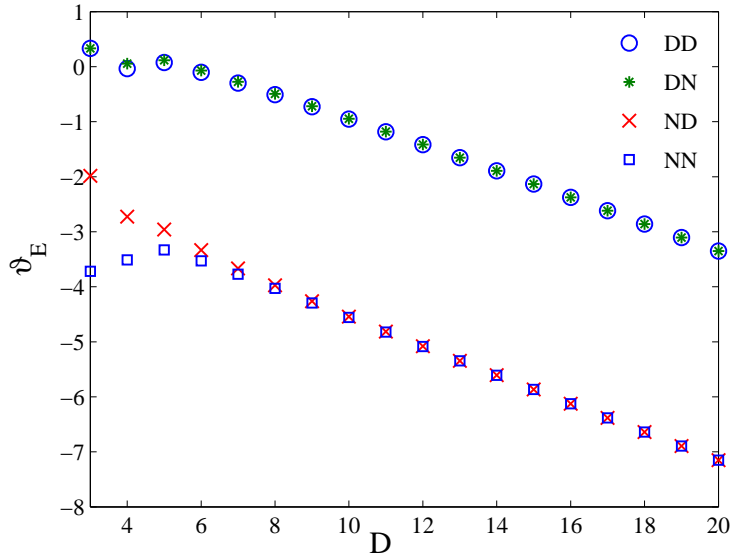} \caption{\label{f1} The dependence of $\vartheta_E$ on dimension $D$.}\end{figure}

In Appendix \ref{a1}, we tabulate the exact and numerical values of $\vartheta_E$ for $3\leq D\leq 12$ in Table \ref{t2}. The dependence of $\vartheta_E$  on $D$ is shown graphically in Fig. \ref{f1}.

From \eqref{eq12_5_12}, it is easy to see that when $D$ is large, the leading order term of $\vartheta_E$ is linear in $D$ with negative coefficient. More precisely,
$$\vartheta_E\sim -\frac{D}{4}\quad \text{for}\, D\gg 1.$$
Hence, when $D$ is large enough, $\vartheta_E$ will become negative, which signifies that proximity force approximation overestimates the Casimir interaction force. Moreover, the magnitude of $\vartheta_E$ will become large when $D$ becomes large. Hence, proximity force approximation becomes less accurate in higher dimensions.

In fact,   from Fig. \ref{f1}, we find that $\vartheta_E$ is negative for all $D\geq 6$, and $\vartheta_E^{\text{ND}}$ and $\vartheta_E^{\text{NN}}$ are negative for all $D\geq 3$. For large $D$, it is obvious from \eqref{eq12_5_12} that $$\vartheta_E^{\text{DD}}\sim\vartheta_E^{\text{DN}},\hspace{1cm}\vartheta_E^{\text{ND}}\sim\vartheta_E^{\text{NN}}.$$
Moreover, observe that
$$\vartheta_E^{\text{DD}}-\vartheta_E^{\text{NN}}=\frac{4(D-1)}{D}\frac{\zeta(D-1)}{\zeta(D+1)}.$$This difference comes from the difference between $\mathcal{J}^{\text{D}}$ and $\mathcal{J}^{\text{N}}$ in \eqref{eq12_6_1}. Hence, in Fig. \ref{f1}, we see that when $D$ is large $\vartheta_E^{\text{DD}}$ and $\vartheta_E^{\text{NN}}$ are approximately two parallel lines with slope $-1/4$ which are vertically $4$ units apart. 

\section{Conclusion}
Using the machinery developed in \cite{7}, we have computed the TGTG formula for the Casimir interaction energy between a sphere and a plate in $(D+1)$-dimensional Minkowski spacetime. We consider massless scalar field with Dirichlet or Neumann boundary conditions. To obtain the formula, we have computed the matrix that changes the spherical wave basis to the plane wave basis, and the matrix that changes the plane wave basis to the spherical wave basis. These results might be useful for other applications.

In $D$-dimensional space, spherical waves are characterized by $(D-1)$ wave numbers $l, m_2,\ldots, m_{D-1}$. Using orthogonality of Gegenbauer polynomials, we find that the TGTG matrix is diagonal in $m_2, m_3,\ldots, m_{D-1}$, and the matrix elements only depend on the two wave numbers $l$ and $m_2$. Hence, the formula for the Casimir interaction energy is not much complicated than the $D=3$ case, except for the appearance of a polynomial of degree $(D-3)$ in $m_2$. Therefore, the formula we derive is useful for both analytical and numerical studies. 

To illustrate the analytical analysis of the formula, we compute the large separation and small separation asymptotic formulas. To compute the large separation leading behavior, we only need   to compute a few matrix elements.     We find that the leading term is proportional to $L^{-D+1}$ if Dirichlet boundary condition is imposed on the sphere, and proportional to $L^{-D-1}$ if Neumann boundary condition is imposed on the sphere. Thus, the former case give rise to stronger Casimir force at large separation.

For the small separation asymptotic   behavior, one has to take into account the contribution from all the matrix elements. Using perturbation method developed in \cite{24}, we obtain the leading order and next-to-leading order terms of the Casimir interaction energy for DD, DN, ND and NN boundary conditions. It is found that for ND and NN boundary conditions, the next-to-leading order term always have sign opposite to the leading order term. For DD and DN boundary conditions, the sign of the leading order and next-to-leading order terms are also opposite of each other when $D\geq 6$. In these cases, the leading order term, which coincides with the proximity force approximation, overestimates the magnitude of the Casimir force. Another observation is that the magnitude of the ratio of the next-to-leading order term to the leading order term grows linearly with dimension $D$, which signifies a larger correction to proximity force approximation in higher dimensions.

The present work is the first step to study the Casimir interaction between two objects of nontrivial geometry in higher dimensional spacetime. In the future, it will be interesting to extend this work to other geometric configurations as well as to other types of quantum fields. 

\begin{acknowledgments}\noindent
  This work is supported by the Ministry of Higher Education of Malaysia  under   FRGS grant FRGS/1/2013/ST02/UNIM/02/2.
\end{acknowledgments}

\appendix
\section{Tabulation of constants}\label{a1}

\begin{table} [h]\caption{\label{t1}The constants $\alpha_D$ and $\beta_D$ for  $3\leq D\leq 12$}

\begin{tabular}{||c|c|c|c|c|| }
\hline
\hline
$D$& $\alpha_D^{\text{DD}}=-\alpha_D^{\text{DN}}$ & $\alpha_D^{\text{NN}}=-\alpha_D^{\text{ND}}$ & $\beta_D^{\text{DD}}=\beta_D^{\text{NN}}$ & $\beta_D^{\text{DN}}=\beta_D^{\text{ND}}$   \\
\hline
&&&&\\
$3$ & $\displaystyle  -\frac{1}{8\pi}$ & $\displaystyle   -\frac{17}{96\pi}$ &$\displaystyle  -\frac{\pi^3}{1440} $ &$\displaystyle  \frac{7\pi^3}{11520}$   \\
&&&&\\
4 & $\displaystyle -\frac{1}{32} $ & $\displaystyle   -\frac{31}{1024} $ &$\displaystyle   -\frac{3\sqrt{2}\zeta(5)}{512}$ &$\displaystyle  \frac{45\sqrt{2}\zeta(5)}{8192} $    \\
&&&&\\
5 & $\displaystyle  -\frac{1}{16\pi} $ & $\displaystyle  -\frac{49}{960\pi}  $ &$\displaystyle  -\frac{\pi^5}{90720}$ &$\displaystyle \frac{31\pi^5}{2903040} $    \\
&&&&\\
6 & $\displaystyle -\frac{3}{256} $ & $\displaystyle  -\frac{71}{8192} $ &$\displaystyle -\frac{15\sqrt{2}\zeta(7)}{16384} $ &$\displaystyle \frac{945\sqrt{2}\zeta(7)}{1048576} $    \\
&&&&\\
7 & $\displaystyle  -\frac{1}{48\pi} $ & $\displaystyle -\frac{97}{6720\pi}  $ &$\displaystyle   -\frac{\pi^7}{6048000}$ &$\displaystyle \frac{127\pi^7}{774144000} $    \\
&&&&\\
8 & $\displaystyle   -\frac{15}{4096}$ & $\displaystyle   -\frac{635}{262144} $ &$\displaystyle -\frac{35\sqrt{2}\zeta(9)}{262144} $ &$\displaystyle  \frac{8925\sqrt{2}\zeta(9)}{67108864}$    \\
&&&&\\
9 & $\displaystyle  -\frac{1}{160\pi} $ & $\displaystyle   -\frac{23}{5760\pi}$ &$\displaystyle  -\frac{\pi^9}{419126400} $ &$\displaystyle \frac{73\pi^9}{30656102400} $    \\
&&&&\\
10 & $\displaystyle  -\frac{35}{32768} $ & $\displaystyle   -\frac{1393}{2097152}$ &$\displaystyle  -\frac{315\sqrt{2}\zeta(11)}{16777216 }$ &$\displaystyle  \frac{322245\sqrt{2}\zeta(11)}{17179869184} $    \\
&&&&\\
11 & $\displaystyle   -\frac{1}{560\pi}$ & $\displaystyle   -\frac{241}{221760\pi}$ &$\displaystyle  -\frac{691\pi^{11}}{20595871296000} $ &$\displaystyle \frac{1414477\pi^{11}}{42180344414208000} $    \\
&&&&\\
12 & $\displaystyle  -\frac{315}{1048576} $ & $\displaystyle -\frac{6027}{33554432}   $ &$\displaystyle  -\frac{693\sqrt{2}\zeta(13)}{268435456} $ &$\displaystyle \frac{2837835\sqrt{2}\zeta(13)}{1099511627776} $    \\
&&&&\\
\hline
\hline
\end{tabular}

\end{table}

\vfill\pagebreak
\begin{table} [h]\caption{\label{t2}The values of $\vartheta_E$ for $3\leq D\leq 12$}

\begin{tabular}{||c|c|c|c|c|c|c|c|c|| }
\hline\hline
\multirow{2}{*}{$D$}& \multicolumn{2}{c|}{$\vartheta_E^{\text{DD}}$ }& \multicolumn{2}{c|}{$\vartheta_E^{\text{DN}}$} &\multicolumn{2}{c|}{$\vartheta_E^{\text{ND}}$} &\multicolumn{2}{c||}{$\vartheta_E^{\text{NN}}$ }  \\
\cline{2-9}
& exact & numerical & exact & numerical & exact & numerical & exact & numerical \\
\hline
&&&&&&&&\\
$3$ & $\displaystyle \frac{1}{3} $ & $\displaystyle   0.3333$ &$\displaystyle \frac{1}{3} $ &$\displaystyle 0.3333 $ & $\displaystyle  \frac{1}{3} - \frac{160}{7\pi^2}$ & $\displaystyle  -1.9826 $ &$\displaystyle  \frac{1}{3} - \frac{40}{\pi^2}$ &$\displaystyle -3.7195  $  \\
&&&&&&&&\\
4 & $\displaystyle   \frac{5}{12} - \frac{7\zeta(3)}{18\zeta(5)}$ & $\displaystyle -0.0342  $ &$\displaystyle \frac{5}{12} - \frac{14\zeta(3)}{45\zeta(5)} $ &$\displaystyle  0.0560 $   & $\displaystyle \frac{5}{12} - \frac{122\zeta(3)}{45\zeta(5)}$ & $\displaystyle   -2.7262$ &$\displaystyle \frac{5}{12} - \frac{61\zeta(3)}{18\zeta(5)} $ &$\displaystyle -3.5119 $  \\
&&&&&&&&\\
5 & $\displaystyle  \frac{1}{2} - \frac{21}{5\pi^2} $ & $\displaystyle  0.0745  $ &$\displaystyle  \frac{1}{2} - \frac{588}{155\pi^2} $ &$\displaystyle 0.1156 $    & $\displaystyle \frac{1}{2} - \frac{ 5292}{155\pi^2} $ & $\displaystyle  -2.9593 $ &$\displaystyle \frac{1}{2} - \frac{ 189}{5\pi^2} $ &$\displaystyle -3.3299 $ \\
&&&&&&&&\\
6 & $\displaystyle   \frac{7}{12} - \frac{2\zeta(5)}{3\zeta(7)}$ & $\displaystyle  -0.1022 $ &$\displaystyle  \frac{7}{12} - \frac{40\zeta(5)}{63\zeta(7)}$ &$\displaystyle  -0.0696 $   & $\displaystyle \frac{7}{12} - \frac{80\zeta(5)}{21\zeta(7)}$ & $\displaystyle  -3.3342$ &$\displaystyle \frac{7}{12} - \frac{4\zeta(5)}{\zeta(7)} $ &$\displaystyle -3.5300 $  \\
&&&&&&&&\\
7 & $\displaystyle   \frac{2}{3} - \frac{200}{21\pi^2}$ & $\displaystyle  -0.2983  $ &$\displaystyle \frac{2}{3} - \frac{ 24800}{2667\pi^2} $ &$\displaystyle -0.2755 $    & $\displaystyle  \frac{2}{3} - \frac{ 114080}{2667\pi^2}$ & $\displaystyle -3.6673   $ &$\displaystyle   \frac{2}{3} - \frac{920}{21\pi^2}$ &$\displaystyle -3.7722 $ \\
&&&&&&&&\\
8 & $\displaystyle  \frac{3}{4} - \frac{5\zeta(7)}{4\zeta(9)} $ & $\displaystyle  -0.5079  $ &$\displaystyle \frac{3}{4} - \frac{21\zeta(7)}{17\zeta(9)} $ &$\displaystyle -0.4931 $    & $\displaystyle \frac{3}{4} - \frac{399\zeta(7)}{85\zeta(9)}$ & $\displaystyle  -3.9738 $ &$\displaystyle\frac{3}{4} - \frac{19\zeta(7)}{4\zeta(9)}  $ &$\displaystyle -4.0301 $ \\
&&&&&&&&\\
9 & $\displaystyle  \frac{5}{6} - \frac{77}{5\pi^2}$ & $\displaystyle -0.7270   $ &$\displaystyle  \frac{5}{6} - \frac{5588}{365\pi^2} $ &$\displaystyle  -0.7179$     & $\displaystyle \frac{5}{6} - \frac{ 128524}{2555\pi^2}$ & $\displaystyle   -4.2634 $ &$\displaystyle  \frac{5}{6} - \frac{253}{5\pi^2}$ &$\displaystyle -4.2935 $\\
&&&&&&&&\\
10 & $\displaystyle  \frac{11}{12} - \frac{28\zeta(9)}{15\zeta(11)}$ & $\displaystyle   -0.9528 $ &$\displaystyle \frac{11}{12} - \frac{1904\zeta(9)}{1023\zeta(11)} $ &$\displaystyle-0.9473  $     & $\displaystyle \frac{11}{12} - \frac{5576\zeta(9)}{1023\zeta(11)}$ & $\displaystyle -4.5422   $ &$\displaystyle  \frac{11}{12} - \frac{82\zeta(9)}{15\zeta(11)}$ &$\displaystyle   -4.5583 $\\
&&&&&&&&\\
11 & $\displaystyle   1 - \frac{163800}{7601\pi^2}$  &$\displaystyle  -1.1835 $ & $\displaystyle  1 - \frac{334807200}{15559247\pi^2} $&$\displaystyle -1.1803 $   & $\displaystyle 1 - \frac{ 892819200}{15559247\pi^2}$ & $\displaystyle  -4.8140$ &$\displaystyle 1 - \frac{436800}{7601\pi^2} $ &$\displaystyle  -4.8225$  \\
&&&&&&&&\\
12 & $\displaystyle  \frac{13}{12} - \frac{5\zeta(11)}{2\zeta(13)} $ & $\displaystyle  -1.4176  $ &$\displaystyle \frac{13}{12} - \frac{682\zeta(11)}{273\zeta(13)}  $ &$\displaystyle -1.4158 $    & $\displaystyle \frac{13}{12} - \frac{25234\zeta(11)}{4095\zeta(13)}$ & $\displaystyle  -5.0811 $ &$\displaystyle \frac{13}{12} - \frac{37\zeta(11)}{6\zeta(13)} $ &$\displaystyle   -5.0856$ \\
&&&&&&&&\\
\hline
\hline
\end{tabular}

\end{table}


\begin{thebibliography}{10}
\bibitem{12} There are a large number of works on this topic.
\end{thebibliography}
\end{document}